# Two-step splitting the expandable graphite for few-layer graphene

Kaiming Liao, Wangfeng Ding, Yuyuan Qin, Zhaoguo Li, Taishi Chen, Jianguo Wan, Fengqi Song*, Min Han*, Guanghou Wang, Jianfeng Zhou

Few-layer graphene sheets are prepared by splitting the expanded graphites using a high-power sonication. Atomic-level quantitative scanning transmission electron microscopy (Q-STEM) is employed to carry out the efficient layer statisticsm, enabling global optimization of the experimental conditions. A two-step splitting mechanism is thus revealed, in which the mean layer number was firstly reduced to less than 20 by heating to 1100°C and then tuned to the few-layer region by a 5-minute $10^4$W/litre sonication. Raman spectroscopic analysis confirms the above mechanism and demonstrates that the sheets are largely free of defects and oxides.

High-yield production of few-layer graphene (FLG) samples with reliable electronic properties forms a critical task in approaching the extensive engineering of graphene, on which two main approaches are highlighted[1-10]. In a "synthesis" approach, e.g. the high-temperature molecular beam epitaxy fabricates the ideally crystalline FLG membranes, whereas the flakes' size are limited for contact bonding[1]. Chemical vapor deposition (CVD) succeeds in the macro-size FLG samples, even free-standing, while the sheets' carrier mobility still needs further improvements[2,3]. The other precursor-based "splitting" approach employs the highly-crystalline precursors including highly oriented pyrolytic graphite (HOPG), carbon nanotube (CNT) and expanded graphite (EG), often produces FLG samples with better electronic properties since shearing of the precursors origins from the imperfection and the crystalline flakes may thus survive the external impact[4-6]. For example, mechanical exfoliation of HOPG results in the FLG films with the best quality[4], leading to the rise of graphene, while the FLG sheets are of very small fraction and often mixed with huge flakes with many layers. The Dai group cracks the multiwall CNTs and the EG powders to prepare the state-of-art FLG nanoribbons[7,8], while formation of large FLG sheets and layer optimization are not emphasized. Such samples provide good transportation quality and enough lateral sizes for contact fabrication[4-8], demonstrating reliable carrier mobility.

Here we present a facile route to fabricate the FLG sheets in a high yield by high-power sonication of the heated expandable graphite (i.e. EG). Firstly, as the precursors, the EG powders exhibit a long-distance crystalline order in the lateral dimension as in the graphite and an enlarged layer distance along the c-axis as in the graphite oxides. A low-temperature and high-vacuum exfoliation can easily split the EG[9]. Li et al.[7] has fabricated ultrasmooth FLG nanoribbons with widths ranging from 10 to 50nm by splitting the EG powders. Reliable conductivities are demonstrated. The same configuration may also apply in the preparation of FLG sheets, while no condition optimization and the related study of its mechanism is reported. Secondly, here the high-power pulsed sonication is firstly reported in the FLG preparation. Hernandez et al.[10] ever reported the synthesis of graphene consisted mostly of mono- and/or few-layer sheets by long low-power sonication of graphite. The present "contratry" conditions may offer an intense and instantaneous micro impacts onto the precursors, leading to the FLG detachment and simultaneously preventing from further non-crystalline process due to prolonged sonication. The FLG sheets can also be prepared. Thirdly to note, the layer number of as-obtained FLG films are counted by an atomic-level calibrated scanning transmission electron microscopy (STEM). Layer counting by atomic force microscopy are hardly applicable in the statistics of many FLG samples even with the aid of optical contrast[6]. High resolution transmission electron microscopy (HR-TEM) provides an efficient searching of the sample[9], while we have to rely on the fringes along the edge of the graphite flakes. The fringes could be invisible particularly for some flakes with very few layers. We show below that the discrete number of graphene layers enables an accurate calibration of STEM intensity with single atomic layer sensitivity[11]. This provides an efficient statistics of the layer numbers and allows the extensive study of the splitting procedure. A two-step procedure is thus concluded to interpret the FLG formation.

All reagents were analytically pure, purchased from Sigma and Nanjing Chemical Reagent Company and used without further purification. The sonication was carried out by JY92-2D Ultrasonic cell Disruptor. The prepared FLG samples were analyzed in a Tecnai F20 transmission electron microscope with a field emission gun operated at 200 kV. The present synthesis involves three key steps: (1) high temperature expanding of the expandable graphite, and (2) ultra-high-power pulsed sonicating of the resulting EG into the FLG sheets and (3) centrifugation (6000rpm for 3min). For the thermal treatments, the raw material of the expandable graphite (Fig. 1a) was transferred into a heating chamber, which allows the thermal treatments up to 1100 °C. The thermal treatment normally takes 1-2 minutes in the present work. It was carried out in a vacuum of $10^{-3}$ pa with a flux of $H_2$ of 0.35 sccm. The expansion of the graphite powder was then observed, resulting in the loose EG powder as shown in Fig.1b. The next step is to disperse 1mg of EG in 10ml of the PmPV/DCE suspension (i.e. the solution of 10mg poly(m-phenylenevinylene-co-2,5-dioctoxy-p-phenylenevinylene) in 100mL 1,2-dichloroethane)for high-power sonication (Fig.1d). The sonic power was set from 100W to 1000 W, with the solution temperature at 40°C for 5 minutes. We then obtained a black suspension with a homogeneous phase of FLG sheets and some visible aggregates (see Fig.1d(1)). The visible flakes could be removed by further centrifugation. Finally, a homogeneous gray dispersion is collected for further characterizations. (see Fig. 1d(2)) Drop-casting the FLG dispersion onto holey TEM grids, the FLG flakes will be suspended on the grids. Fig. 1e displays an overview of a typical FLG sample, taken by STEM imaging, where the FLG flakes are highlighted by the light blue shade. The lateral sizes of the flakes are a few microns. All the flakes present good crystalline condition as demonstrated by the electron diffraction. Considering the crystalline condition of the precursors, the previous report on EG-based FLG preparation [Dai's work] and the Raman measurement (see below), we believe our samples are of good electronic properties.

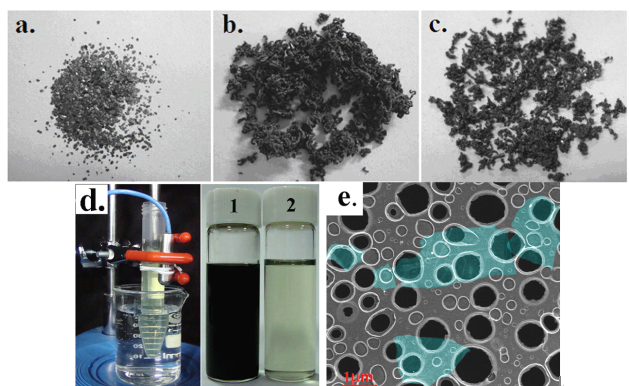

**Fig.1** The photos of a. expandable graphite flakes. b. resultant expanded at 1100°C. c. resultant expanded at 750°C. Note the distinct volume increase at 1100°C. d. the configuration in which the EG was dispersed in pre-prepared PmPV/DCE mixed solution for sonication. d-1. the black suspension after sonicating at 600W for 5min; d-2. the solution with the FLG sheets after centrifugation. e. the STEM image of the as-prepared FLG sheets on the holey grids. The FLG sheets are highlighted by blue shades.

We determine the layer number of the FLG sheets by a calibrated STEM, where the STEM intensity is contributed by a high angle angular dark field (HAADF) detector. Essentially, we demonstrate a calibration curve of the STEM intensity against the FLG layer number. As widely known, HR-TEM fringes (dark lines) may appear along the curved or folded edge of the FLG flakes. The number of the dark lines determines the layer number of the FLG flake.[12-15] The validity of this method has been cross-checked using Raman spectrum,[12,16] nanobeam electron diffraction[14] and electron energy loss spectroscopy.[12] We then image the flake by STEM (Fig. 2a) and obtain the HAADF contribution (Fig. 2c) of the layer number as determined by the HR-TEM (Fig. 2b). This finally makes a point in the calibration curve in Fig 2d. A typical edge is shown in Fig. 2b, where 7 lines are found. This leads to a layer number of 7 with the error of 1, concerning the influence of Fresnel fringes. As it is folded in Fig. 2a, 14 layers correspond to the HAADF contribution obtained in the line profile (Fig. 2c). Fig. 2d shows the HAADF intensity as a function of graphene layer number, where a quasilinear relationship is apparent. Therefore, the layer number of the FLG flakes can be obtained by direct analysis of the STEM intensity, which provides an efficient statistics of the layer numbers and allows the extensive study of the splitting procedure. The measurements then give the layer number of 3-5 for the FLG flakes in Fig. 1e.

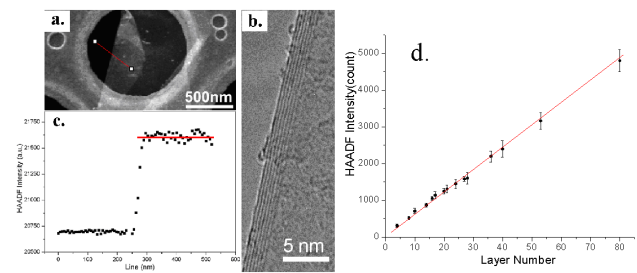

**Fig.2** Layer counting.(a) is an STEM image of a FLG flake with folded edge, (b) is the HR-TEM image of the edge, and (c) is the line profile marked by the red line in (a). The dark lines indicate the number of layers of 7 in (b), while 14 layers correspond to 800 HAADF intensities due to the folded edge. The calibration curve. (d) The HAADF intensities are plotted against the layer numbers. The data and the error bars are taken from the mean value and the corresponding standard deviation from the selected FLG sheets. A linear fitting is valid particularly for small layer numbers.

Layer statistics are carried out to optimize the experimental conditions such as the expanding temperature and sonicating power/period manipulating the layer numbers of the FLG flakes. Firstly, we obtained a series of products via changing the ultrasonic power when the heating temperature is 1100°C and the sonication period is 5min. With increasing the power of ultrasonic irradiation from 200W to 600W, one finds that the layer numbers of the FLG flakes largely decreased and the fraction of the FLG sheets increased obviously as shown in Fig.3a. Nevertheless, there is no further improvement if we further increase the power to 900W. Secondly, we also investigated the products prepared at different sonic periods with the same conditions that the heating temperature is 1100°C and the sonic power is fixed to 600W. The layer numbers were found to increase with the prolonging of the irradiation durations ( see Fig.3b), which should be ascribed to the fact that the FLG flakes with very few layers are decomposed due to sustained high power irradiation. Finally, in the previous reports[7, 17,18], the high-temperature expanding was regard as a critical step for the FLG preparation in the sonochemistry route. We studied the samples of the EG powders expanded at different temperatures and confirmed the expanding temperature is a critical issue. For the examples as shown in Fig. 4d, we carried out the expansion at 750 °C. The layer numbers of the FLG flakes were even more than 20 layers. The mean layer numbers kept large even if we boosted the sonic power, while the layer numbers of the product reached less than 10 even at the small sonication power of 200W by using the sample expanded at 1100 °C (see Fig. 4b).It is even clearer by seeing the contrast in their STEM images (Fig. 4a and c). The result suggests the sufficient expansion of the expandable graphite occurs at 1100 °C, which can also be shown from the volume of expanded samples (see Fig. 1b and c). We thus obtained the optimized conditions for the production of FLG sheets with the smaller layer numbers and high FLG yields.

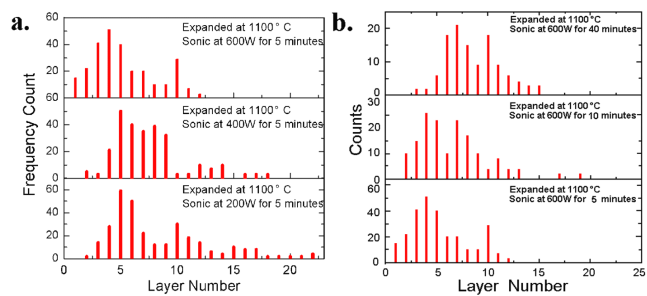

**Fig.3** Statistics of layer numbers in optimizing the sonication parameters. The parameters are marked in the figure. (a) shows the change of sonication powers, and (b) shows the change of sonication durations.

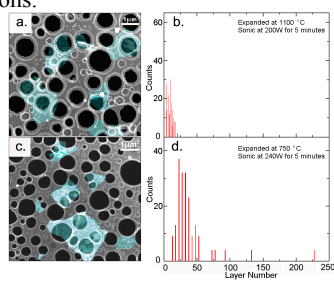

**Fig.4** Layer statistics of the FLG sheets with different expending temperatures as marked in the figure (b) and (d); and (a) and (c) show their corresponding STEM images, respectively.

Raman spectroscopy confirms the two-step splitting mechanism and demonstrates the quality of the sheets. The spectra were collected by a JY HR800 and a NT-MDT Nanofinder 30 machine using the 488nm excitation laser. Three features are considered, in which the "G" peak around 1600cm$^{-1}$ refers to the graphitic ordering, the "D" band around 1350cm$^{-1}$ refers to the disordered fractions and the "2D" peak around 2700cm$^{-1}$ checks the FLG sheets [19,20]. Fig. 5a and b shows the Raman spectra of EG powder and the powder after expanding at 1100 °C in a vacuum of 10$^{-3}$ pa with a flux of H$_2$ of 0.35 sccm, respectively. The G peaks in both spectra are sharp and almost identical in the position and the peak width, indicating the good graphitic crystalline conditions. Note that the small D band of the EG powder almost disappears after the first step of the high temperature treatment, which shows the high temperature treatment "cure" the disordered fractions and the defects. Fig. 5c is the Raman spectrum of a FLG sheet and Fig. 5e shows its Raman spectrum expanded to 3000 cm$^{-1}$. We can see clear 2D peak as previously reported for the FLG sheets [20] Its broadening as compared to the 2D peak of single layer graphene indicates the layer numbers are around 5-7 layers, agreeing to the STEM measurement and the atomic force microscopic measurement. Therefore, we can conclude an initial step of expanding and H$_2$ warm curing and a second step of sonic splitting are implemented in the present preparation.

The quality of as-prepared FLG sheets can also be assessed. Electron diffraction shows the six-fold pattern as stated above. Raman spectrum in Fig 5c also demonstrates the FLG sheets are free of obvious disordered compositions since even the nanosized graphene sheets presents good contribution to the "D" band [21] The suppression of the "D" band and the appearance of "2D" peak also confirms the good quality of the FLG sheets. The EG is of much less organic functional groups than the precursors of graphite oxide, hence one may infer the resultant FLG sheets' quality is pre-determined by their raw materials as our experimental results favor to support the report of Bourlinos et al[22]. We compare the Raman spectrum of the FLG flakes prepared from oxide graphite [23](see Fig.5d). It is easily found that there is a relatively high D band and a much broadened G peak, indicating mass defect content.

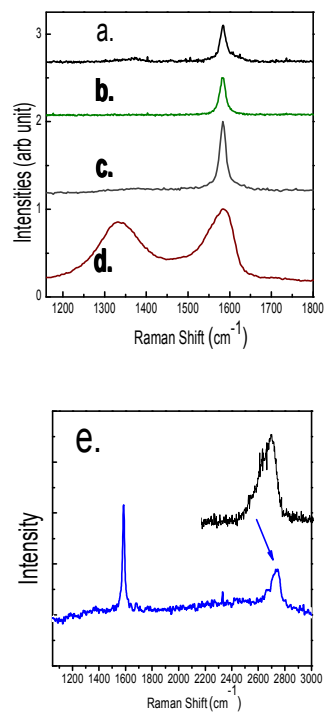

**Fig.5** Raman spectrum of the dispersed fraction in the solid state: a. the EG powder; b. the high temperature treatmented EG powder; c. the as-prepared FLG sheets; d. the FLG flakes prepared from graphite oxide; e. the full range Raman spectrum of as-prepared FLG sheets.

The FLG sheets are formed by a two-step splitting mechanism, in which thermal expansion enlarges the layer distance of the EG powder and the high-power sonication splits the loose EG for separate FLG flakes. The van der Waals forces between the graphene layers decreased rapidly with increasing the layer spacings. The sufficient expansion leads to a double layer spacing, which makes the EG easily split. On the critical temperature, Schniepp et al.[18] reported the synthesis of functionalized single graphene sheets by rapid preheating graphite oxide to 1050 °C. Again, Li et al.[7] reported the successful synthesis of graphene nanoribbons by preheated expandable graphite to 1000°C. They believe that rapid heating of the expandable graphite caused violent formation of volatile gaseous species from the intercalates, leading a loose stack of FLG sheets. Our experiments support the above theory with a critical temperature of 1100 °C for the sufficient expansion. The second splitting is then to detach the FLG flakes by dense sonication. After the previous expansion, the obtained samples are only partially exfoliated and still contain disperse domains of staked graphitic layers. Therefore, it could be still energy-consuming for full overcoming the van der Waals forces between the graphene layers, on which several approaches have been developed. Lv et al.[9] found that the ultrahigh vacuum exfoliation may achieve the detachment even at a low-temperature. Janowska et al.[24] who considered the high microwave absorption properties of the graphite fabricated the FLG sheets in aqueous solution of ammonia by a microwave irradiation. Li et al.[7] reported long sonication in special solution can fabricate the graphene nanoribbons. Hernandez et al.[10] also reported long sonication leads to the complete layer detachment. Herein, we find the FLG sheets could be rapidly fabricated by ultra-high-power sonication, and the layer numbers of products could be regulated via changing the sonication power. It is interesting to note that long sonication will destroy the FLG flakes. The result suggests that the attachment between graphene layers should present a critical energy. High-power and instantaneous sonication provides a unique micro-splitting process on the basis of the acoustic cavitations. The formation, growth and implosive collapse of bubbles in the liquid generates many localized hot-spots with transient temperatures of 5000~25,000 K and transient pressures of ~1800 atm[25], where the heating and cooling rates may be in excess of $10^{10}$ K s$^{-1}$.[26] Such strong impact from ultrahigh sonication may surpass and then break the connection directly. Instantaneous and dense sonication leads to rapid preparation of the FLG flakes.

In summary, the FLG sheets with regulated layer number have been successfully prepared and the optimized conditions are obtained. Considerable monolayer graphene sheets can be found. Raman spectroscopic analysis of these sheets suggests the flakes to be largely free of defects and oxides.A two step splitting process is suggested to explain the FLG formation, in which sufficient expansion at enough high temperature and layer detachment by dense and instantaneous sonication both play important roles. This approach enables an alternative route toward mass production of graphene sheets and this may be meaningful for further fundamental studies and future applications of the FLG sheets.

We thank the National Natural Science Foundation of China (Grant numbers: 90606002, 11075076, and 10775070) and the National Key Projects for Basic Research of China (Grant numbers: 2009CB930501, 2010CB923401) for supporting this project. the Program for New Century Excellent Talents in University of China Grant No. NCET-07-0422. The useful discussions with Dr. Yanbin Chen, Prof. Xiaoning Zhao and Mr. Qianjin Wang were also acknowledged.

## Notes and references

*National Lab of Solid State Microstructures, Nanjing University, 210093, Nanjing, P. R. China Fax: 02583595535 Tel:02583686745*
*E-mail:songfengqi@nju.edu.cn, sjhanmin@nju.edu.cn*

Electronic Supplementary Information (ESI) available: [details of structure characterizations]. See DOI: 10.1039/b000000x/